\documentclass[a4paper,10pt,twocolumn,preprint,3p]{elsarticle}
\usepackage{geometry,mathenv,epsfig,graphicx,amsmath,amssymb,mathtext,amsthm,textcomp,subfig,float}
\usepackage{epstopdf}
\usepackage{lineno,hyperref}
\usepackage{color}
\usepackage{array}
\modulolinenumbers[5]

\begin{document}

\begin{frontmatter}

\title{Plasma acceleration limitations due to betatron radiation}


\author[1]{V.Shpakov\corref{cor1}}
\cortext[cor1]{Corresponding author}
\ead{vladimir.shpakov@lnf.infn.it}

\author[1]{E.Chiadroni}
\author[1]{A.Curcio}
\author[1,3]{H.Fares}
\author[1]{M.Ferrario}
\author[1]{A.Marocchino}
\author[1,4]{F.Mira}
\author[2]{V.Petrillo}
\author[2]{A.R.Rossi}
\author[1]{S.Romeo}

\address[1]{INFN Laboratori Nazionali di Frascati, Via Enrico Fermi 40, 00044 Frascati (Rome), Italy}
\address[2]{INFN-Milan and Department of Physics, University of Milan, Via Celoria 16, 20133 Milan, Italy}
\address[3]{Department of Physics, Faculty of Science, Assiut University, Assiut 71516, Egypt}
\address[4]{Department of Basic and Applied Sciences for Engineering (SBAI) and INFN-Roma1}

\begin{abstract}
High energy spread caused by the longitudinal size of the beam is well known in wake-field acceleration. Usually this issue can be solved with beam loading effect that allows to keep accelerating field nearly constant, along the whole duration of the beam. In this work, however, we would like to address another source of energy spread that arises at high energy, due to betatron radiation.
\end{abstract}

\begin{keyword}
PWFA, betatron radiation, energy spread
\end{keyword}

\end{frontmatter}

\section{Introduction}
The reason of high attention towards the plasma wake-field acceleration is high longitudinal electric field, that can be induced by the driver (laser or particle beam) inside the plasma,which potentially, can be useful for a new generation of high energy ($\sim TeV$ level) colliders. In addition to the longitudinal field, however, the beam inside the plasma experiences also the transverse electric field, which leads to oscillation of the particles inside the ion cavity and, as a consequence, emission of radiation that we know as betatron radiation (BR). Power of BR ($P_{br}$) is rising along with the energy of the beam \cite{Esarey2002}:
\begin{equation}
P_{br}\approx r_em_ec^3\gamma^2k_p^2r_\beta^2/3,
\label{eq1}
\end{equation}
where $r_e$ is the classical electron radius, $m_e$ is the electron mass, $c$ is the speed of light, $\gamma$ is the Lorentz factor of the particle, $k_p$ is the plasma wave-vector, $r_\beta$ is the amplitude of oscillations. Since the amount of energy that we pump into the beam ($P_{wf}$) remains constant \cite{Rosenzweig1988}:
\begin{equation}
P_{wf}\approx c^2m_e\omega_p,
\label{eq1.1}
\end{equation}
where $\omega_p$  is the plasma frequency, it is reasonable to assume that there is a limit to plasma-based acceleration \cite{Zimmerman,Barletta}, when $P_{wf}=P_{br}$. In addition to this question we also address the issues of cooling \cite{Deng} of the beam and energy spread generation due to the BR.
 
\section{Betatron radiation and methods}

Moving inside the ion cavity behind the driver, electrons of the witness beam will oscillate under the influence of the transverse electric field (Fig.\ref{fig1}). The emitted BR has some resemblances to synchrotron or wiggler radiation and, at high energy (depending on plasma density it can be from $\sim100~GeV$ for $n_p=10^{19}~cm^{-3}$, up to $\sim1~TeV$ for $n_p=10^{16}~cm^{-3}$) of the beam, the power lost in BR appears comparable with the power of the plasma wake-field. Thus several effects should be taken into account when approaching high energy: 
\begin{figure}[h]
    \includegraphics[width=0.49\textwidth]{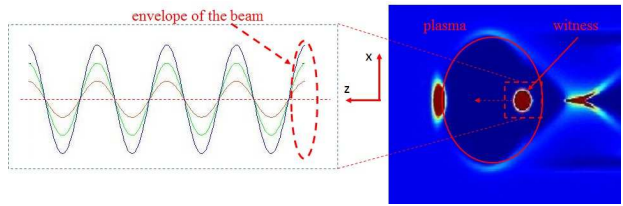}
    \caption{In the simplest case the amplitude depends on the particle position at the entrance to the plasma, but it always stays within the envelope of the beam.}
    \label{fig1}
\end{figure}
losses due to BR and the cooling of the beam, for instance. In addition, unlike plane wigglers, the amplitude of oscillations inside the ion cavity can be different for different electrons (see Fig.\ref{fig1}). As a consequence different electrons will loose different amount of energy and thus we will have generation of the energy spread caused by BR.
In order to take into account all the effects mentioned above we can employ the equations of motion for a single electron:
\begin{align}
\frac{dp_z}{dt}&=F_{z,p}-F_{z,rr}
\label{eq2}\\
 \frac{dp_x}{dt}&=F_{x,p}-F_{x,rr},
\label{eq3}
\end{align}
where $p$ is the momentum of the electron, $F_{p}$  is the force created by the plasma, and $F_{rr}$ is the radiation reaction force. Both forces will be derived from the assumption that maximum longitudinal electric field is in the wave-breaking limit \cite{Rosenzweig1988,Esarey2009}, i.e.
\begin{equation}
E_{z,max}=\frac{m_ec\omega_p}{e},
\label{eq4}
\end{equation}
and transverse electric field is \cite{Jackson}
\begin{equation}
E_{x}=\frac{k_p^2m_ec^2}{2e}x,
\label{eq5}
\end{equation}
where $x$ is the transverse coordinate of the particle with respect to the axis of the bubble. The radiation reaction force components can be calculated under the following assumptions:
\begin{equation}
F_{z,rr}\approx F_{rr}=-\frac{e^2}{6\pi\epsilon_0m_e^2c^4}\left(\frac{dp_\mu}{d\tau}\right)^2,
\label{eq6}
\end{equation}
for the longitudinal component, and transversal components are:
\begin{equation}
F_{x,rr}=\frac{e^2}{6\pi\epsilon_0m_e^2c^4}\left(\frac{d^2p_\mu}{d\tau}-\frac{p_\mu}{m_e^2c^2}\left(\frac{dp_\mu}{d\tau}\right)^2\right),
\label{eq7}
\end{equation}
where $\epsilon_0$ is the electric constant, and $\tau$ is the proper time of the particle. Using Eqs.~(\ref{eq4}-\ref{eq7}) in (\ref{eq2}) and (\ref{eq3}) we can arrive to the following system of equations:
\begin{align}
&\frac{d^2x}{dz^2}+\left(\frac{k_p^2}{\pi\gamma} z+\frac{e^2}{12\pi\epsilon_0m_ec^2} k_p^2\right) \frac{dx}{dz}+\frac{k_p^2}{2\gamma} x=0
\label{eq8}\\
&\frac{d\gamma}{dz}=\frac{k_p^2}{\pi}z_0-\frac{e^2}{24\pi\epsilon_0m_ec^2}k_p^4\gamma^4x^2,
\label{eq9}
\end{align}
where $z_0$ is the relative position of the witness with respect to the driver. By solving these equations we can track down the evolution of the beam parameters (e.g. transverse size, emittance, energy, etc...). For more detailed  derivation of these equation we will refer to \cite{Deng,Deng2}.

\section{Results of calculations}
Equations (\ref{eq8}) and (\ref{eq9}) have been integrated by using Runge-Kutta method. To investigate the limits of wake-field acceleration described in the introduction we have chosen the initial energy $1~GeV$ (energy spread $0.01\%$), which is correspond to the 1st iteration target energy for the plasma-driven photo injector of EuPRAXIA project \cite{EUPraxia1,EUPraxia2}. The transverse size of the beam was always considered to matched size of the beam \cite{Ferrario2012}:
\begin{equation}
\sigma_{x}=\sqrt[4]{\frac{2}{\gamma}}\sqrt{\frac{\epsilon}{k_p}},
\end{equation}
where $\epsilon$ is the normalized beam emittance. In this work we considered kilometers of plasma, that requires staging, however, possible effects from staging itself were not included. Following parameters of the beam and plasma were considered: 
\begin{table}[h]
\begin{center}
\begin{tabular}{ c c c c }
\textnumero&$\epsilon, ~mm\cdot mrad$ & $n_p,~cm^{-3}$ & line on figures\\
1)&$1.0$ & $1.0\times10^{16}$ & \textcolor{blue}{blue, solid}\\
2)&$0.5$ & $1.0\times10^{16}$ & \textcolor{red}{red, dashed}\\
3)&$1.0$ & $0.5\times10^{16}$ & \textcolor{green}{green, dotted}
\end{tabular}
\end{center}
\end{table}

\subsection{Energy and energy spread}
Evolution of the energy was linear at the start, with slight deviation at higher energy (see Fig.~\ref{fig6}). The energy spread can be generated by two effects. First represented by the $1^{st}$ element on right side of Eq.~\ref{eq9}, caused by the longitudinal size of the beam. Second, represented by the $2^{nd}$ element on the right side of Eq.~\ref{eq9}, caused by BR emission. To isolate the contribution due to the BR, we assume a perfect beam loading, which results in $1^{st}$ element being equal to all particles, while $2^{nd}$ element depends on the transverse position of the particle.
\begin{figure}[h]
    \includegraphics[width=0.49\textwidth]{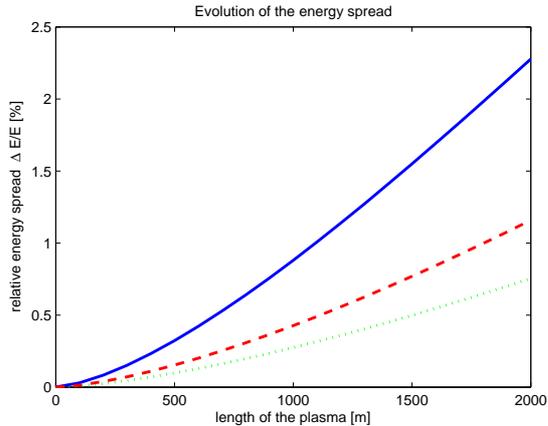}
    \caption{Energy spread growth during PWFA. In cases 1) and 2) energy at the end was $\sim25~TeV$, in case 3) $\sim19~TeV$ (see Fig.\ref{fig6}).}
    \label{fig3}
\end{figure} 
By the time when electron beam reaches the energy around $25~TeV$ the BR generates energy spread at the level of $\sim2.2\%$ (Fig.\ref{fig3}, solid line). Decrease of the original emittance helps to slowdown energy spread growth (Fig.~\ref{fig3}, dashed line). More effective way to delay growth of the energy spread is to use a lower plasma density (Fig.~\ref{fig3}, dotted line), but this causes a decrease of the accelerating gradient.

\subsection{Size and emittance of the beam}
\begin{figure}[h]
    \includegraphics[width=0.49\textwidth]{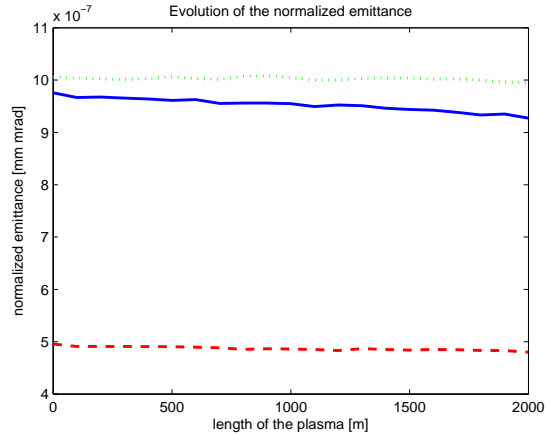}
    \caption{Evolution of the emittance after $2~km$ of plasma. In 1st case (solid line) the emittance has changed from $0.97$ to $0.92 \mu m$, which is $\sim5\%$. For the cases 2 and 3 the decrease of the emittance is even less.}
    \label{fig4}
\end{figure}

The behavior of the transverse size of the beam was almost identical for all three cases. Increasing the energy the beam is rapidly focused at the beginning of acceleration, with slower focusing at higher energy. For the case 1) matched transverse size changed from $\sim1.3\mu m$  to $\sim0.1\mu m$ at the end as expected. The cooling, however, appears to be insignificant compared to the acceleration length. For $1~mm\cdot mrad$ beam and $n_p=1.0\times10^{16}~cm^3$ plasma density, after 2 km of acceleration (this is only plasma) the emittance decreased for only $\sim5\%$ (see Fig.~\ref{fig4}, solid line). In cases of lower plasma density or emittance, the cooling was even less significant.
\subsection{Limit of acceleration}
\begin{figure}[h]
    \includegraphics[width=0.49\textwidth]{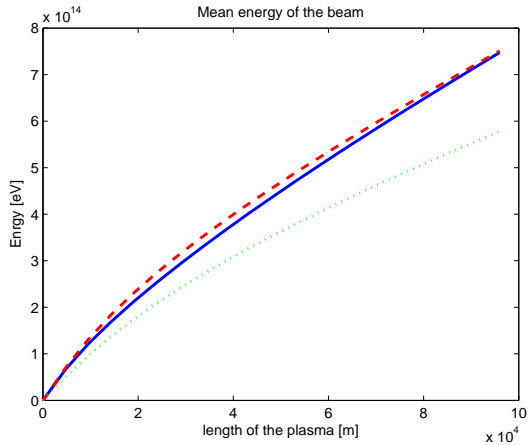}
    \caption{Evolution of the beam energy. With increase of the energy BR losses go up as well, however, it never fully negates the acceleration.}
    \label{fig6}
\end{figure}

\begin{figure}[h]
    \includegraphics[width=0.49\textwidth]{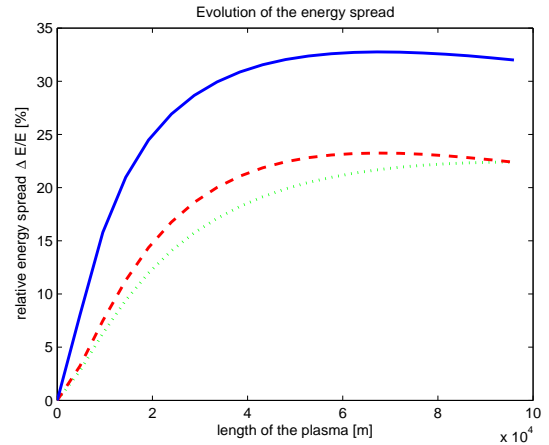}
    \caption{Energy spread growth depends on plasma and beam parameters, but eventually reaches maximum at $\sim20-30\%$.}
    \label{fig7}
\end{figure}

Finally we tried to find the maximum energy that can be achieved with plasma-based accelerators. To do so we integrated Eqs.(\ref{eq8},\ref{eq9}) for $\sim100~km$. Our calculations have shown that limit does not exist (Fig.\ref{fig6}) and, apparently, BR power never reaches $100\%$ the power of PWFA, although it does reduce the overall accelerating gradient.
 The same (or similar) result was achieved in \cite{Deng,Deng2}. At the same time, however, energy spread growth due to the BR becomes a significant issue. Despite of the fact that it reaches saturation, it stays at the level of $20-30~\%$ (see Fig.\ref{fig7}).
As expected, the reduction of the emittance or the plasma density also decreases the velocity of the energy spread growth, but does not solve this issue completely.

\section{Conclusion}

The main goal of this work was to establish whether plasma acceleration limit exists. From purely theoretical point of view, according to our calculations, it does not exist. Despite of the fact that with rising energy BR losses go up as well, they never reach $100~\%$ of PWFA gradient and acceleration can continue infinitely, but with lower effective gradient. From practical point of view, however, under assumption that we are aiming to $\leq1~\%$ energy spread beam, there is a limit dictated by the energy spread growth due to the BR. For the parameters that were used in this work ($n_p\approx10^{16}~cm^{-3}$ and emittance $\sim1~mm\cdot mrad$) at $\sim15~TeV$ of energy spread already reaches  $1\%$ or more, thus limiting maximum achievable energy.

Regarding the cooling of the beam by means of BR. The fact that we did not found any limits to plasma acceleration can be explained by the presence of an equilibrium between BR cooling and acceleration. Higher energy leads to increase of the BR power, which results in faster beam cooling and, as a result, in smaller beam size, which, in its turn, decreases the BR power and increases the effective gradient, leading to the higher energy. Thus, apparently, beam cooling can not be simply disregarded. At the same time it is difficult to use BR cooling for practical applications due to its very low efficiency ($\sim5\%$ of emittance decrease after $2~km$ of plasma, see Fig.~\ref{fig4}).

\subsection{Important notes.}
It is important to highlight that limits found in this work strongly depend on parameters of the beam and plasma itself. In general, the lower plasma density will result in higher achievable energy and slower energy spread growth. The same can be said about the smaller emittance of the beam, although decreasing the emittance is less effective in this regard. 

Another important point is the fact that this calculations do not take into account any possible degradation of the beam emittance during PWFA, which is certainly the case according to the full scale simulations \cite{Esarey2009,Gessner2012,Architect3}. During our study we added BR effects into the Architect \cite{Architect1,Architect2} code. Since in order to see any BR effects it is necessary to simulate kilometers of plasma our simulations were inconclusive and are not included in this work. However, it is important to underline that in case of deteriorating emittance energy spread growth appeared to be much more severe.
\section*{Acknowledgment}
This work was supported by the European Union‘s Horizon 2020 research and innovation programme under grant agreement No. 653782. One of the authors, H. Fares, would like to acknowledge support from Academy of Scientific Research and Technology (ASRT) in Egypt and INFN in Italy (ASRT-INFN joint project).

\section*{References}

\bibliography{mybibfile}

\end{document}